\documentclass[aps,prb,twocolumn,showpacsamsmath,preprintnumbers,amsmath,amssymb]{revtex4-2}

\usepackage{graphicx}% Include figure files

\usepackage{color}

\usepackage{bbm}
\usepackage{bm}% bold math
\usepackage{color}
\usepackage{dcolumn}% Align table columns on decimal point
\usepackage{epstopdf}
\usepackage{graphicx}% Include figure files
\usepackage{lipsum}
\usepackage{float}
\usepackage{amsmath}
\usepackage{subfigure}
\usepackage{gensymb}
\usepackage{ulem}

\usepackage[hidelinks]{hyperref}
\hypersetup{
    colorlinks,
    citecolor=blue,
    filecolor=blue,
    linkcolor=blue,
    urlcolor=blue
}

\newcommand{\be}{\begin{equation}}
\newcommand{\ee}{\end{equation}}
\newcommand{\bea}{\begin{eqnarray}}
\newcommand{\eea}{\end{eqnarray}}

\newcommand{\s}{\sigma}

\newcommand{\up}{\uparrow}
\newcommand{\down}{\downarrow}

\newcommand{\trace}{\mathop{\mathrm{Tr}}\nolimits}
\newcommand{\ie}{i.\,e. }

\begin{document}

%\begin{frontmatter}
%\title{Signatures of Nagaoka's ferromagnetism in transport through a quadruple quantum dot attached to ferromagnetic leads}
\title{Spin-polarized transport in  quadruple quantum dots attached to ferromagnetic leads}

\author{Piotr Trocha}
\email{ptrocha@amu.edu.pl}
\author{Emil Siuda}
\author{Ireneusz Weymann}
\address{Institute of Spintronics and Quantum Information, Faculty of Physics,
	Adam Mickiewicz University, 61-614 Pozna{\'n}, Poland}

%\date{\today}

\begin{abstract}
Motivated by the experimental evidence of the Nagaoka ferromagnetism
in quantum dot systems by Dehollain { \it et al}. \cite{Dehollain},
we search for possible confirmation of such kind of ferromagnetism
by analyzing the spin-resolved transport properties of a quadruple quantum dot system
focusing on the linear response regime.
In particular, we consider four quantum dots arranged in a two-by-two square lattice,
coupled to external ferromagnetic source and drain electrodes.
Turning on and off the specific conditions for the Nagaoka ferromagnetism to occur by changing
the value of the intra-dot Coulomb interactions, we determine the transport coefficients, including the linear conductance,
tunnel magnetoresistance and current spin polarization.
We show that a sign change of the current spin polarization may be an indication
of a ferromagnetic order of Nagaoka type which develops in the system.
\end{abstract}

%\begin{keyword}
	%quantum dot, spin valve, Nagaoka ferromagnetism, spin-polarized transport
%\end{keyword}

\maketitle

%%%%%%%%%%%%%%%%%%%%%%%%%%%%%%%%%%%%%%%%%%%%%%%%%%%%%%
%%%%%%%%%%%%%%%%%%%%%%%%%%%%%%%%%%%%%%%%%%%%%%%%%%%%%%
\section{Introduction}
%%%%%%%%%%%%%%%%%%%%%%%%%%%%%%%%%%%%%%%%%%%%%%%%%%%%%%
%%%%%%%%%%%%%%%%%%%%%%%%%%%%%%%%%%%%%%%%%%%%%%%%%%%%%%

More than half a century ago Nagaoka predicted theoretically
the existence of a form of ferromagnetism,
which does not arise naturally in any material~\cite{Nagaoka}.
Nagaoka explained rigorously that materials with one caveat can become magnetic.
However, experimental evidence of such ferromagnetic order
has been elusive for decades until very recent achievements in experimental techniques,
which allowed to observe the Nagaoka ferromagnetism in
the system comprised of coupled quantum dots (QQs) \cite{Dehollain}.
In particular, the experimental setup consisted of four QDs arranged in a two-dimensional lattice,
for which the specific conditions required for the Nagaoka ferromagnetism to occur
were obtained by tuning the gate voltages to the three electron occupation regime.
Furthermore, using a very sensitive electric sensor, it was possible
to determine the spin orientation of the electrons and convert it into a measurable electric signal,
which enabled the confirmation of whether or not the electron spins align ferromagnetically.
A general theoretical analysis of a small number of quantum dots focusing
on the Nagaoka ferromagnetism has been recently performed by Buterakos {\it et al.} \cite{Buterakos}.
Although four quantum dot system arranged in square-lattice structure
is a minimal model to observe the Nagaoka ferromagnetism, we note that a partially ferromagnetic state
can also emerge in a triangular quantum dot \cite{oguri}.

Here, motivated by the recent experimental evidence \cite{Dehollain},
we investigate the transport properties of a four quantum dot system
attached for ferromagnetic leads, searching for signatures of
Nagaoka ferromagnetism in spin-resolved transport characteristics.
Using the Green's function formalism,
we determine the linear conductance, tunnel magnetoresistance (TMR)
and the spin polarization of the current flowing through the system.
We show that the spin polarization exhibits a sign change when
the system crosses over from antiferromagnetic to ferromagnetic alignment.
This change of sign may be thus an indirect fingerprint
of a ferromagnetic order of three spins occupying the quadruple QD system,
which is associated with the Nagaoka's prediction.

We point out that the spin-polarized transport properties of quadruple quantum dots system are rather unexplored.
Although a perfect spin-polarization induced by Coulomb interactions has been reported theoretically \cite{kaganJMMM},
most of existing considerations focus on unpolarized transport of quadruple quantum dots \cite{kaganPRB,kaganJETP,li, liu}.
A quadruple quantum dot device in a square-like configuration has been realized and studied experimentally \cite{Thalineau}
and the Kondo effect has also been observed in such a structure \cite{shang}. Moreover,
it has been shown that a quadruple quantum dot system
implements the minimal mechanism for acting as a self-contained quantum refrigerator \cite{Venturelli}.

%%%%%%%%%%%%%%%%%%%%%%%%%%%%%%%%%%%%%%%%%%%%%%%%%%%%%%
%%%%%%%%%%%%%%%%%%%%%%%%%%%%%%%%%%%%%%%%%%%%%%%%%%%%%%
\section{Theoretical framework}
%%%%%%%%%%%%%%%%%%%%%%%%%%%%%%%%%%%%%%%%%%%%%%%%%%%%%%
%%%%%%%%%%%%%%%%%%%%%%%%%%%%%%%%%%%%%%%%%%%%%%%%%%%%%%

\begin{figure}[t]
	\begin{center}
		\includegraphics[width=0.45\textwidth,angle=0]{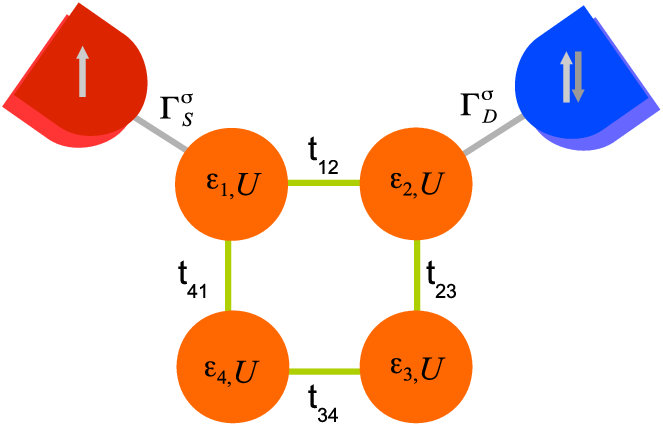}
		\caption{ \label{Fig:1} The schematic illustration of
			a $2\times2$ quantum dot lattice coupled to external ferromagnetic leads,
			whose magnetic moments can form either parallel or antiparallel configuration.}
	\end{center}
\end{figure}

%%%%%%%%%%%%%%%%%%%%%%%%%%%%%%%%%%%%%%%%%%%%%%%%%%%%%%
%%%%%%%%%%%%%%%%%%%%%%%%%%%%%%%%%%%%%%%%%%%%%%%%%%%%%%
\subsection{Model Hamiltonian}
%%%%%%%%%%%%%%%%%%%%%%%%%%%%%%%%%%%%%%%%%%%%%%%%%%%%%%
%%%%%%%%%%%%%%%%%%%%%%%%%%%%%%%%%%%%%%%%%%%%%%%%%%%%%%

The system taken into considerations consists of four QDs
arranged in a square geometry, with two QDs attached to
external ferromagnetic leads, as schematically depicted in Fig.~\ref{Fig:1}.
The magnetic moments of ferromagnetic electrodes
can be aligned either in parallel (P) or antiparallel (AP) configuration.
The system is modeled by the Hamiltonian of the general form
\begin{equation}\label{Eq:1}
H=H_{c}+H_{4QD}+H_{T}.
\end{equation}
The first term, $H_{c}$, describes the spin-polarized electrons
in the ferromagnetic source ($\beta=S$) and drain ($\beta=D$) leads and takes the form
\begin{equation}\label{Eq:2}
H_{\rm c}=\sum_{\beta=S,D}\limits\sum_{{\mathbf
k}\sigma}\limits\varepsilon_{\beta{\mathbf k}\sigma}
     c^\dag_{\beta{\mathbf k}\sigma}c_{\beta{\mathbf k}\sigma}.
\end{equation}
The second term describes the quadruple QD and is given by
\begin{equation}\label{Eq:3}
H_{\rm 4QD}=\sum_{\sigma,i=1}^{4}\limits\varepsilon_{i\sigma}d_{i\sigma}^{\dagger}d_{i\sigma}
-\sum_{\sigma}\sum_{\langle i,j\rangle} t_{ij} d_{i\sigma}^\dagger d_{j\sigma}
+\sum_{i=1}^{4}\limits U_i n_{i\uparrow}n_{i\downarrow},
%+\frac{\Phi}{2}\sum_{\langle i,j\rangle}\sum_{\sigma\sigma'}\limits n_{i\sigma}n_{j\sigma'},
\end{equation}
with $\varepsilon_{i\sigma}$ denoting single-particle energy level of $i$-th
quantum dot, whereas $t_{ij}$ is the hopping amplitude between neighboring quantum dots.
In turn, $U_i$ is the Coulomb repulsion energy on the $i$th QD.
In the following, we assume equal hopping parameters ($t_{ij}=t$)
and the same on-dot Coulomb interactions ($U_i=U$).

The last term in (\ref{Eq:1}) describes the tunnel coupling of QD-1 to
source electrode and QD-2 to drain lead. It is given by
\begin{eqnarray}\label{Eq:4}
H_{\rm T}=\sum_{{\mathbf k}\sigma}V_{S\mathbf{k}\sigma}c_{S\mathbf{k}\sigma}^{\dagger}d_{1\sigma}
+ \sum_{{\mathbf k}\sigma}V_{D\mathbf{k}\sigma}c_{D\mathbf{k}\sigma}^{\dagger}d_{2\sigma}
+ \mathbf{\rm{H.c.}},
\end{eqnarray}
with $V_{\mathbf{k}\beta i}$ being the coupling amplitude of given dot to its lead.
Furthermore, the coupling matrix elements,
$V_{\mathbf{k}\beta i}$, are assumed to be $\textbf{k}$-independent.

The width of the QD's level due to the coupling to the $\beta$th
external electrode can be parametrized by the coupling strength
$\Gamma_{\beta}^{\sigma}=2\pi \langle |V_{\beta\mathbf{k}\sigma}|^2\rangle\rho_{\beta\sigma}=\Gamma_{\beta}(1\pm p_\beta)$,
where $p_\beta$ denotes the spin polarization of the magnetic lead $\beta$
and the upper (lower) sign corresponds to
tunneling electrons with spin parallel (antiparallel) to the leads' magnetic moment.
Above, $\langle|V_{\beta{\bf k}\sigma}|^{2}\rangle$ denotes
the corresponding average over ${\bf k}$ and $\rho_{\beta\sigma}$ stands for
the density of electron states in the metallic lead $\beta$.
Within the wide band approximation the density of states
$\rho_{\beta\sigma}$, and therefore, the coupling strength $\Gamma_{\beta}^{\sigma}$
becomes energy independent, \ie a constant parameter.
In order to distinguish two collinear magnetic configurations of external metallic leads we set
\begin{equation}\label{Eq:5}
\Gamma_{S}^{\sigma}=\Gamma_{S}(1\pm p_S)
\end{equation}
for source ($\beta=S$) electrode, and
\begin{equation}\label{Eq:6}
\Gamma_{D}^{\sigma}=\Gamma_{S}(1\pm sp_D)
\end{equation}
for drain ($\beta=D$) electrode
with $s=1$ for the parallel and $s=-1$ for the antiparallel magnetic configuration.

%%%%%%%%%%%%%%%%%%%%%%%%%%%%%%%%%%%%%%%%%%%%%%%%%%%%%%
%%%%%%%%%%%%%%%%%%%%%%%%%%%%%%%%%%%%%%%%%%%%%%%%%%%%%%
\subsection{Method}
%%%%%%%%%%%%%%%%%%%%%%%%%%%%%%%%%%%%%%%%%%%%%%%%%%%%%%
%%%%%%%%%%%%%%%%%%%%%%%%%%%%%%%%%%%%%%%%%%%%%%%%%%%%%%

To calculate the relevant transport characteristics, including the conductance,
tunnel magnetoresistance and current spin polarization,
we employ the Green's function technique.
Following Meir and Wingreen~\cite{meir}, one
can show that the current flowing through the system is given by
\begin{equation}\label{Eq:7}
J=\frac{e}{h}\sum_{\sigma}\int {\mathrm d}\varepsilon [f_S(\varepsilon)-f_D(\varepsilon)]\mathcal{T}_{\sigma}(\varepsilon),
\end{equation}
where $f_{\beta}(\varepsilon) = \{\exp[(\varepsilon-\mu_\beta)/k_BT] + 1\}^{-1}$
is the Fermi-Dirac distribution function for the lead $\beta$ with $\mu_\beta$ and $T$
denoting the corresponding chemical potential and
temperature, while $k_B$ stands for the Boltzmann constant.
Furthermore, $\mathcal{T}_{\sigma}(\varepsilon)$ is the transmission coefficient
associated with spin-$\sigma$ carriers.
The transmission coefficient can be expressed by
the Fourier transforms of the retarded ($\mathbf{G}_{\sigma}^r$) and advanced
($\mathbf{G}_{\sigma}^a$) Green's functions of the dots and by the coupling
matrices $\mathbf{\Gamma}_{\beta}^{\sigma}$ as
$\mathcal{T}_{\sigma}(\varepsilon)=\trace{[\mathbf{\Gamma}_{S}^{\sigma}\mathbf{G}_{\sigma}^r\mathbf{\Gamma}_{D}^{\sigma}\mathbf{G}_{\sigma}^a]}$.
Above,
$\mathbf{\Gamma}_{S}^{\sigma}=\operatorname{diag}(\Gamma_{S}^{\sigma},0,0,0)$
and $\mathbf{\Gamma}_{D}^{\sigma}=\operatorname{diag}(0,\Gamma_{D}^{\sigma},0,0)$.
The Green's functions have been calculated by the equation of
motion technique within the Hubbard I approximation for
the Coulomb term~\cite{trocha2007}. The relevant correlators,
$n_{ij\sigma}\equiv\langle d_{i\sigma}^{\dag}d_{j\sigma}\rangle$,
have been calculated self-consistently with the help of identities
\begin{equation*}\label{Eq:8}
n_{ij\sigma}=-i\int\frac{\mathrm{d}\varepsilon}{2\pi}G_{ji\sigma}^{<}(\varepsilon),
\end{equation*}
where $G_{ji\sigma}^{<}(\varepsilon)$ denotes the lesser Green's function,
which can be derived by using the Keldysh relation
\begin{eqnarray}
\mathbf{G}^{<}_{\s}=\mathbf{G}^{r}_{\s}\mathbf{\Sigma}^{<}_{\s}\mathbf{G}^{a}_{\s},\label{Eq:4}
\end{eqnarray}
whereas the lesser self-energy can be obtained from the following formula
\begin{eqnarray}
\mathbf{\Sigma}^{<}_{\s}=\mathbf{\Sigma}^{<}_{S\s}+\mathbf{\Sigma}^{<}_{D\s}=
i[f_S(\varepsilon)\mathbf{\Gamma}_{D}^\s+
   f_D(\varepsilon)\mathbf{\Gamma}_{D}^\s],\label{Eq:5}
\end{eqnarray}
which is valid for interactions taken within the mean field approximation.
Note that only $n_{ij\sigma}$ for $i=j$ and for $i,j=i+1$ are needed.
As the four QDs form a closed loop, the index $j$ for $i=4$ reads $j=1$.

%%%%%%%%%%%%%%%%%%%%%%%%%%%%%%%%%%%%%%%%%%%%%%%%%%%%%%
%%%%%%%%%%%%%%%%%%%%%%%%%%%%%%%%%%%%%%%%%%%%%%%%%%%%%%
\section{Numerical results}
%%%%%%%%%%%%%%%%%%%%%%%%%%%%%%%%%%%%%%%%%%%%%%%%%%%%%%
%%%%%%%%%%%%%%%%%%%%%%%%%%%%%%%%%%%%%%%%%%%%%%%%%%%%%%
%

\begin{figure*}
\begin{center}
\includegraphics[width=1.0\textwidth,angle=0]{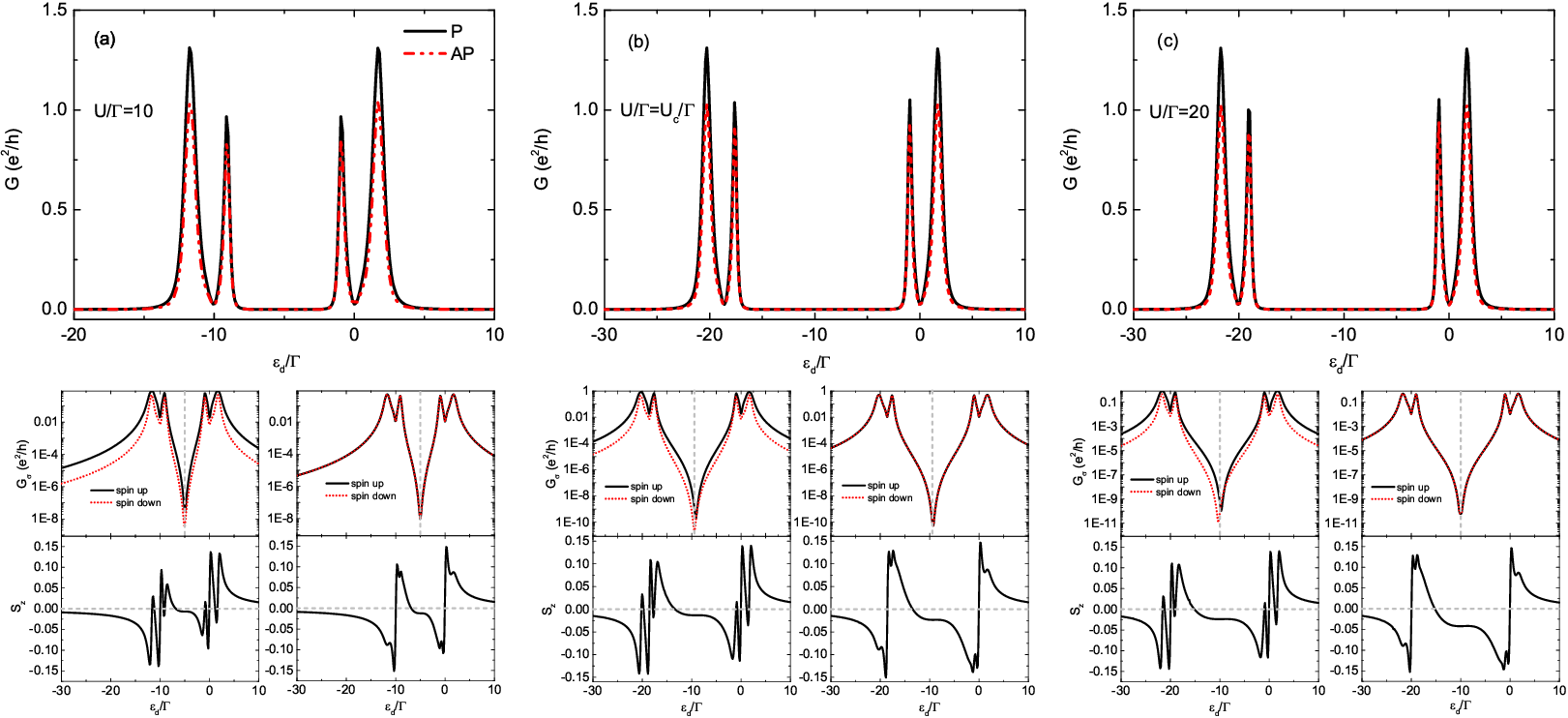}
\caption{\label{Fig:2} The linear conductance as a function of the dot's level position calculated
for indicated values of the Coulomb energy and for the parallel and
antiparallel magnetic configuration. The other parameters are $t=\Gamma$,
$p=0.5$, $k_BT=0.1\Gamma$ and $\Gamma\equiv 1$ used as energy unit.
Insets show spin-resolved conductances in logarithmic scale and average z-component of the
total spin for parallel (left panel) and antiparallel (right panel)
magnetic alignment. For AP configuration spin-up and spin-down conductances are the same, whereas for P configuration
they are different. Moreover, in the former case the positions of the dips
are located exactly at $\varepsilon_d=-U/2$. However, this ceases to be true in the P configuration for
$U/t\geq U_c/t$ when the minima in the spin-resolved conductances move away particle-hole symmetry point.
}
  \end{center}
\end{figure*}
In this section we present the numerical results on spin-polarized
transport in the system under consideration.
For numerical calculations we assume spin degenerate
and equal dot levels, $\varepsilon_{i\up}=\varepsilon_{i\down}\equiv\varepsilon_d$
(for $i = 1, 2, 3, 4$ and $\sigma=\up, \down$).
We also assume symmetrical couplings to ferromagnetic electrodes, $\Gamma_S=\Gamma_D\equiv\Gamma$,
and that both leads are made of the same material, $p_S=p_D\equiv p$.

%Moreover, we relate energy quantities to the dot-lead coupling.
%For the sake of simplicity we also assume the same Coulomb
%parameters for the four dots, $U_{i} = U$ for $i=1,2,3,4$.

Performing the diagonalization of the Hamiltonian of QDs decoupled from external electrodes, cf. Eq.~(\ref{Eq:3}),
we obtained the eigen-energies dependence on the ratio of $U/t$ for a given dot's level position $\varepsilon_d$.
As for the Nagaoka's ferromagnetism the system must be occupied by $N-1$ electrons
($N$ is the number of quantum dots), here we concentrate on the three electron states.
There are states, which differ by spin value,
as three electron states can have spin equal to either $1/2$ or $3/2$.
The most important feature is that for relatively small value of the $U/t$ ratio
the electrons are aligned antiferromagneticaly with the resulting spin equal to $1/2$,
whereas for $U/t$ larger than a certain critical value ($U_c/t$)
the system tends to the ferromagnetic order with spin $3/2$.
For $\varepsilon_d=0$, we found this critical value to be $U_c/t=18.5831$.
Having found the critical value, we have calculated the transport characteristics for three regimes,
when: (i) $U/t<U_c/t$, (ii) $U/t=U_c/t$ and (iii) $U/t>U_c/t$.

We start our investigations by calculating the conductance in the linear response regime
for two magnetic configurations of the external electrodes. In Fig.~\ref{Fig:2}
we show the linear conductance as a function of dot's level position calculated for
indicated values of the Coulomb energy $U$ and for the parallel and
antiparallel magnetic configuration. The conductance reveals
four peaks associated with resonant states. The positions of these peaks
are approximately at $t$, $-\sqrt{3}t$, $t-U$, $-\sqrt{3}t-U$ and are
independent on the magnetic configuration of the system.
Between the two pairs of the peaks a characteristic Coulomb gap emerges,
where the conductance is greatly suppressed.
This gap obviously scales linearly with the Coulomb correlation parameter $U$.
Moreover, the spectrum of resonances is symmetric with respect to
the particle-hole symmetry point $\varepsilon_d=-U/2$.
Interestingly, the conductance between each pair of peaks is
strongly suppressed revealing a characteristic dips.
It vanishes at the dot's level positions
$\varepsilon_d=0$ and $\varepsilon_d=-U$.
This originates from the quantum interference
of electron waves transmitted through two available paths,
which acquire different phases. For these specific points
the phase shift between the two paths is exactly equal to
$\pi$ resulting in a totally destructive quantum interference.

\begin{figure}[t!]
	\begin{center}
		\includegraphics[width=0.45\textwidth,angle=0]{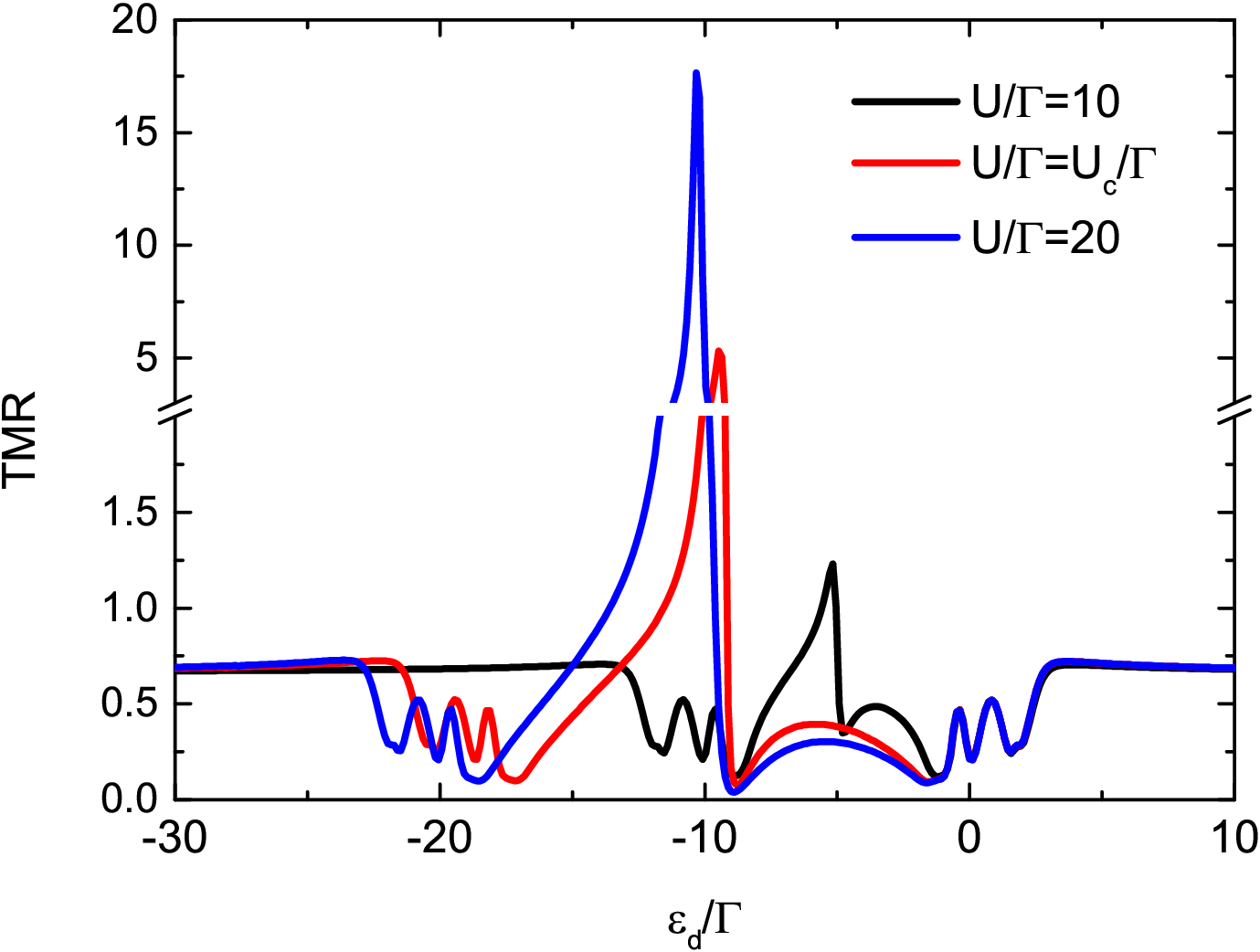}
		\caption{\label{Fig:3} The TMR as a function of the dot's level position calculated
			for the indicated values of the Coulomb correlation.
			The other parameters are the same as in Fig.~\ref{Fig:2}.}
	\end{center}
\end{figure}
The conductance reveals qualitatively the same behavior for the P and
AP configuration, with only quantitative differences. For a better
understanding of this difference let us consider the tunneling magnetoresistance,
which is defined as
\begin{equation}\label{TMR}
  \mathrm{TMR}=\frac{G_P-G_{AP}}{G_{AP}}
\end{equation}
with $G_P$ ($G_{AP}$) denoting the conductance in the P (AP)
alignment. The presented in Fig.~\ref{Fig:3} TMR reflects nontrivial differences
between the two magnetic configurations.
When the dots’ levels are well above or well below the Fermi level of the leads,
the TMR tends to the Julliere’s value for magnetic tunnel junctions~\cite{Julliere},
\ie $\mathrm{TMR}=2p^2/(1-p^2)$, which gives $\mathrm{TMR}=2/3$ for $p=0.5$.
In turn, when the dot levels/one of the resonances crossed the Fermi energy,
the situation becomes more complex and the TMR gets suppressed
below the Julliere's value. It is worth noting that a similar suppression of
the TMR was also found in single quantum dots ~\cite{weymann}.
However, in the present case the suppression of TMR
is stronger than that in a single dot and reveals characteristic oscillations around the resonances
in the linear conductance.
More interestingly, the TMR becomes strongly enhanced in the vicinity
of the particle-hole symmetry point (\ie $\varepsilon_d=-U/2$),
where the corresponding conductance, in both magnetic configurations,
is strongly suppressed due to the Coulomb blockade effect.
This is a consequence of the fact that the suppression
of conductance in the Coulomb gap strongly depends on the magnetic
configuration of the system.
A similar enhancement of TMR was found in double QDs system,
where peaks in the TMR were present not only in the vicinity of the particle-hole symmetry
point but also near antiresonances~\cite{trocha2007}.
However, in the latter case the mechanism leading to these peaks was rather different.

\begin{figure}
	\begin{center}
		\includegraphics[width=0.45\textwidth,angle=0]{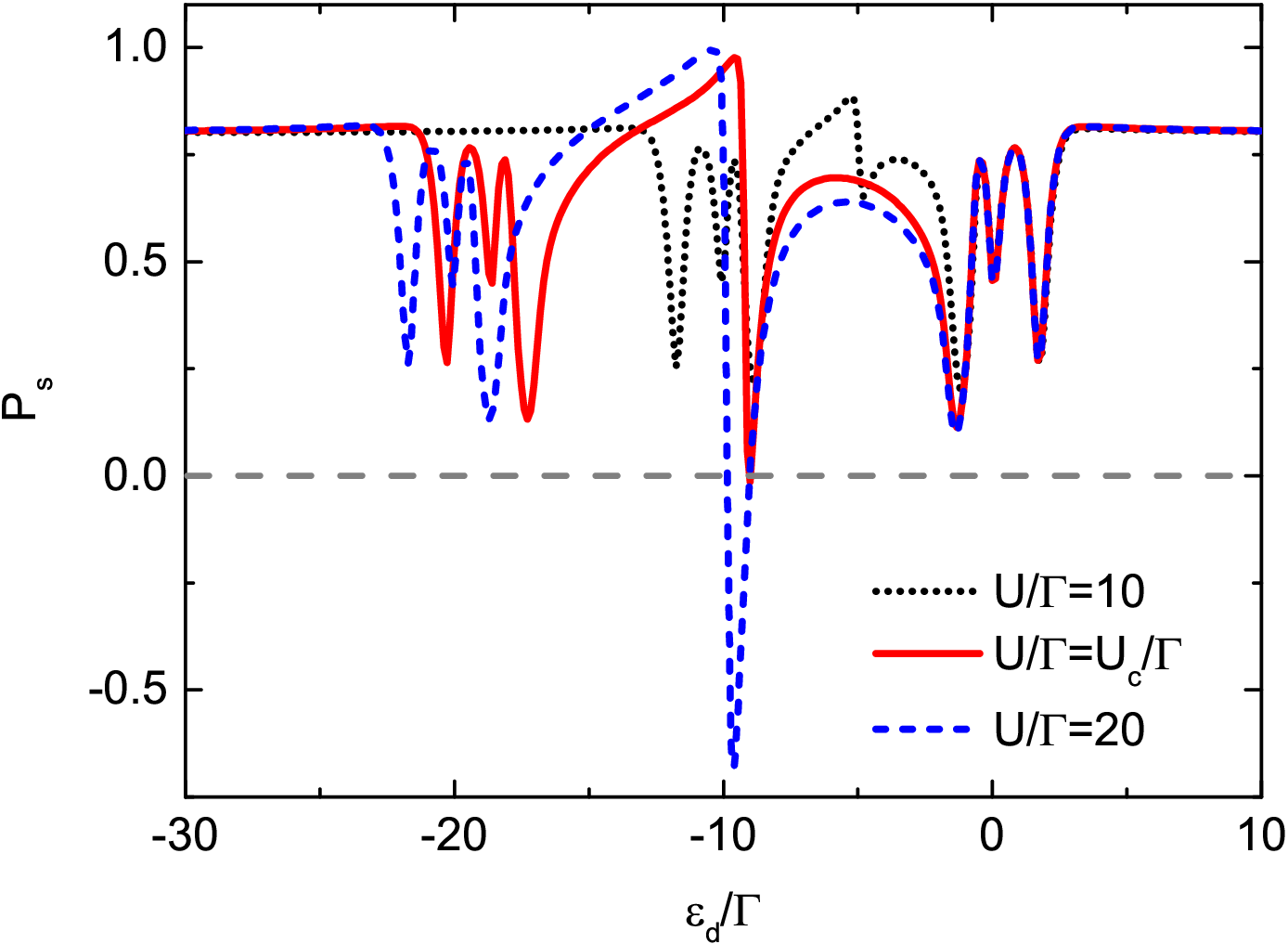}
		\caption{\label{Fig:4} The current spin polarization as a function
			of the dot's level position calculated
			for indicated values of $U$.
			The other parameters are as in Fig.~\ref{Fig:2}.}
	\end{center}
\end{figure}

Finally, we have calculated the current spin polarization, which is defined as
\begin{equation}\label{Ps}
  \mathrm{P}_S=\frac{G_{\up}-G_{\down}}{G_{\up}+G_{\down}},
\end{equation}
where $G_{\sigma}$ is linear conductance in spin $\sigma$ ($\sigma=\up , \do$) channel.
As the spin polarization in the AP configuration vanishes due to
the assumed symmetry of the system ($\Gamma_S=\Gamma_D$ and $p_S=p_D$),
we only consider the parallel alignment.
In Fig.~\ref{Fig:4} the spin polarization dependence on
the dot level position is plotted for indicated values of the Coulomb correlation parameter.
When the dots’ levels are well above or well below the Fermi level of the leads,
the spin polarization achieves relatively high values, $P_S\gtrsim 0.75$.
Consequently, in the considered system one can obtain larger current spin polarization
than the spin polarization of used magnetic leads, \ie $P_S>p$.
In turn, when the dot levels cross the Fermi energy, $P_S$ reveals
oscillations around the value of $0.5$ with alternating enhancement and
suppression of the current spin polarization. This behavior seems to be
independent of the strength of Coulomb parameter.

The most striking feature of $P_S$ appears in the vicinity
of particle-hole symmetry point. When crossing $\varepsilon_d=-U/2$,
the current spin polarization changes significantly its value
especially for relatively high Coulomb parameter and can even change sign.
For both $U=U_c$ and $U>U_c$, the current becomes perfectly spin polarized
for certain dot levels, \ie $P_S\approx 1$ for $\varepsilon_d\lesssim -U/2$.
However, for $\varepsilon_d\gtrsim -U/2$ it can be perfectly suppressed
when $U=U_c$ ($P_S\approx 0$) or can change sign when $U>Uc$ ($P_S<0$).
This result seems to be indirectly related to the Nagaoka's ferromagnetism,
although more studies are needed in this direction. Especially, one should
show how the current spin polarization scales with coupling to external leads
$\Gamma$. Here, we only show the case with $t=\Gamma$ but our preliminary
results (not shown) suggest that there exists non-trivial dependence on
couplings.

%%%%%%%%%%%%%%%%%%%%%%%%%%%%%%%%%%%%%%%%%%%%%%%%%%%%%%
%%%%%%%%%%%%%%%%%%%%%%%%%%%%%%%%%%%%%%%%%%%%%%%%%%%%%%
\section{Conclusions}
%%%%%%%%%%%%%%%%%%%%%%%%%%%%%%%%%%%%%%%%%%%%%%%%%%%%%%
%%%%%%%%%%%%%%%%%%%%%%%%%%%%%%%%%%%%%%%%%%%%%%%%%%%%%%

In conclusion, we have studied the spin-resolved transport properties
of a quandruple quantum dot system attached to ferromagnetic contacts.
Using the Green's function technique, we have determined the linear conductance,
TMR and current spin polarization. We have shown how these quantities
change when the condition for the Nagaoka ferromagnetism becomes
satisfied. Furthermore, we have shown that the current spin polarization
changes sign when the system exhibits ferromagnetic order of Nagaoka's type.
Our results indicate that studying the spin-resolved transport properties
can give further insight into nontrivial magnetic order associated
with Nagaoka ferromagnetism.

%%%%%%%%%%%%%%%%%%%%%%%%%%%%%%%%%%%%%%%%%%%%%%%%%%%%%%
%%%%%%%%%%%%%%%%%%%%%%%%%%%%%%%%%%%%%%%%%%%%%%%%%%%%%%
\section*{Acknowledgments}
This work was supported by the National Science Centre
in Poland through the Project No. 2017/27/B/ST3/00621.
E. S. acknowledges support of the National Science Center in Poland
through the research Project No. 2018/31/D/ST3/03965.

%%%%%%%%%%%%%%%%%%%%%%%%%%%%%%%%%%%%%%%%%%%%%%%%%%%%%%
%%%%%%%%%%%%%%%%%%%%%%%%%%%%%%%%%%%%%%%%%%%%%%%%%%%%%%

\end{document}